\documentclass[a4paper,11pt]{article}
\usepackage{amsmath,amsthm,latexsym,amssymb,amsfonts,epsfig}
\usepackage{fontenc,dsfont,layout}
\usepackage{hyperref}
\usepackage{psfrag}
\usepackage[cp1250]{inputenc}
\usepackage{graphics}
\numberwithin{equation}{section}

\DeclareMathAlphabet{\mathpzc}{OT1}{pzc}{m}{it}

\begin{document}

\title{Classical bouncing Universes from vector fields}
\author{Micha{\l} Artymowski\thanks{Michal.Artymowski@fuw.edu.pl} $\;\,$  and $\;$
           Zygmunt Lalak\thanks{Zygmunt.Lalak@fuw.edu.pl}}
\date{\it  Institute of Theoretical Physics, Faculty of Physics, University of Warsaw ul. Ho\.{z}a 69, 00-681 Warszawa, Poland} 
\maketitle

\begin{abstract}
For the anisotropic Universe filled with massless vector field in the General Relativity frame we obtain bouncing solution for one of scale factors. We obtain the Universe with finite maximal energy density, finite value of $R,R^{\mu\nu}R_{\mu\nu},R^{\mu\nu\alpha\beta}R_{\mu\nu\alpha\beta}$ and non-zero value of a scale factor for directions transverse to a vector field. Such a bounce can be also obtained for a massive vector field with kinetic initial conditions, which gives isotropic low energy limit. We discuss the existence of a bounce for a massless vector field with additional matter fields, such as cosmological constant or dust. We also discuss bouncing solution for massless vector field domination in $n+2$ dimensional space-time.  
\end{abstract}

\section*{Introduction} \label{sec:wstep}

The evolution of Abelian vector fields in cosmology has been discussed in the context of generating isotropic inflation \cite{Golovnev:2008cf}, large scale structure \cite{Dimopoulos:2006ms,Dimopoulos:2008yv,Himmetoglu:2009mk,Dimopoulos:2009vu,Dimopoulos:2009am}, dark energy and background anisotropies \cite{Dimopoulos:2006ms,Koivisto:2008xf}. One shall note, that unlike scalar fields, we observe elementary vector fields in accelerator experiments. Thus, we expect them to appear in the early Universe, for instance as a gauge bosons. Since then, they are considered to be a viable alternative for scalar fields in solving problems of classical cosmology. As we shall show in this paper, vector fields may also result in a finiteness of the energy density of the Universe and generate bouncing evolution of one of scale factors in the GR frame. 

Our original motivation to study the evolution of a massless vector field in the GR frame was to find the classical limit for its evolution in Loop Quantum Cosmology \cite{Artymowski:2010by}. However, it turned out that the massless vector field domination can generate a bounce already in the GR frame. Let us note, that bouncing solutions were found long before LQC \cite{Bojowald:2006da,Ashtekar:2009vc,Artymowski:2008sc}, for example in $f(R)$ theories \cite{Gurovich}, in the cycling Universe \cite{Steinhardt:2001st} or as a result of quantum corrections \cite{Parker:1973qd}. In the present paper  we study the detailed model mentioned already in the previous paper \cite{Artymowski:2010by}. It turns out, that the model has, several advantages: one does not need to introduce any modification of General Relativity (like in $f(R)$ theory) or to introduce any exotic kinetic terms for fields, which would dominate the Universe during a bounce (like in ekpyrotic$\backslash$cyclic Universe). Solutions, that we have obtained here may form an interesting background for particle physics models of the early Universe. 

In this paper we will use convention $8\pi G=M_{pl}^{2}=1$.

This paper is organised as follows: In section \ref{sec:bezmasowywektor} we discuss the evolution of space-time for a massless vector field domination, for which we find bouncing solution of the scale factor $a(t)$. In section \ref{sec:dominacjaAbezmasnD} we discuss a massless vector field domination in $n+2$ dimensional space-time. In section \ref{sec:BBreal} we introduce more realistic (i.e. with isotropic low energy limit) scenarios with a bounce, such as the domination of a massive vector field with kinetic initial conditions or the Universe filled with a massless vector field, cosmological constant and/or dust. Conclusions are presented in section \ref{sec:concl}.
\\*

\section{Massless vector field domination} \label{sec:bezmasowywektor}

\subsection{Canonical vector field}\label{sec:veccanonical}

The evolution of a subdominant vector field $\mathcal{A}_\mu$ in FRW space-time has been analysed in e.g. \cite{Dimopoulos:2006ms,Himmetoglu:2009mk}. It has been shown, that the homogeneous (background) component of a vector field lies in the space-like part of a four-vector. Then, one may choose one of axes to point in the direction of a background vector field. Let us choose it to be the $z$ axis, which gives
\begin{equation}
\mathcal{A}_\mu=A_\mu(t)+\delta A_\mu(\vec{x},t),\qquad A_{\mu}=A(t)\delta^z_{\ \mu},\qquad\delta A_{\mu}=(\delta A_t,\delta\vec{A}), \label{eq:dekompozycjaAmu}
\end{equation}
where $\delta A_\mu$ is a vector field perturbation, which may contribute to the generation of the large scale structure. In this paper we will consider the Universe filled with a background vector field and a perfect fluid. A vector field breaks the isotropy of space, thus the metric tensor takes the form of 
\begin{equation}
g_{\mu\nu}=\text{Diag}(1,-a(t)^2,-a(t)^2,-b(t)^2)\ , \label{eq:metrykav}
\end{equation}
where $a(t),b(t)$ are scale factors. Then the Einstein equation looks as follows
\begin{eqnarray}
H(H+2\mathcal{H})=\rho=\rho_f+\rho_A\ , \label{eq:Ein00v}\\
\frac{\ddot{a}}{a}+\frac{\ddot{b}}{b}+H\mathcal{H}=-p_\perp=-p_f-p_A\ , \label{eq:Ein11v}\\
2\frac{\ddot{a}}{a}+H^2=-p_\parallel=-p_f+p_A\ , \label{eq:Ein33v}
\end{eqnarray}
where $\rho_f,p_f$ are the energy density and pressure of a perfect fluid and $H=\frac{\dot{a}}{a},\mathcal{H}=\frac{\dot{b}}{b}$ are Hubble parameters along $x,y$ ($H$) and $z$ ($\mathcal{H}$) axis. The vector field's energy density and pressure is defined by
\begin{equation}
\rho_A=\frac{1}{2b^2}\left(\dot{A}^2+m^2A^2\right),\qquad p_A=\frac{1}{2b^2}\left(\dot{A}^2-m^2A^2\right) \label{eq:rhoApA}\ .
\end{equation}
Equations of motion of a vector field and a perfect fluid looks as follow
\begin{equation}
\ddot{A}+(2H-\mathcal{H})\dot{A}+m^2A=0,\qquad\dot{\rho}_f+(2H+\mathcal{H})(\rho_f+p_f)=0\ .\label{eq:ruchuAf}
\end{equation}

\begin{figure}[h]
\includegraphics*[height=5cm]{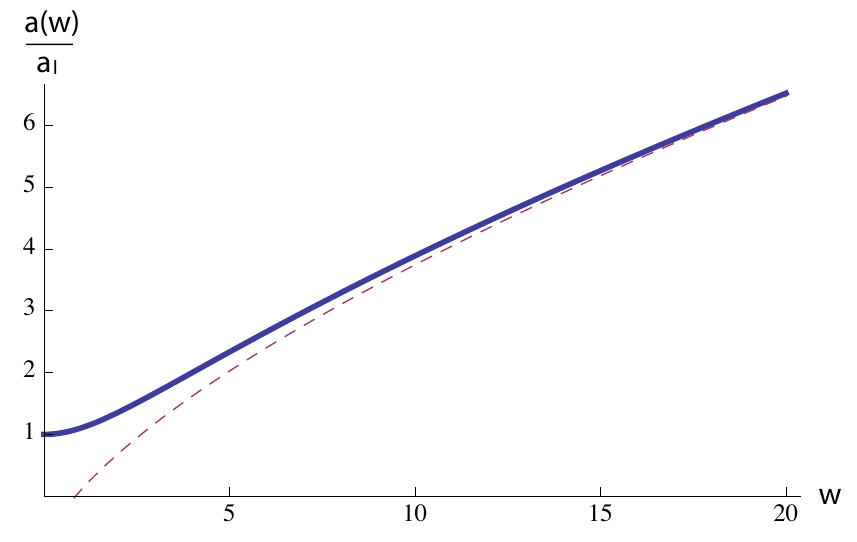}
\hspace{0.3cm}
\includegraphics*[height=5cm]{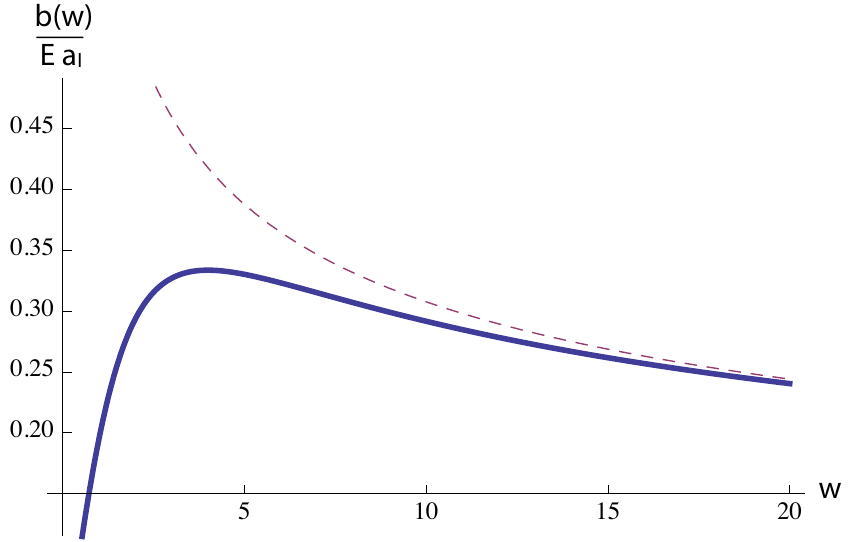}
\caption{Left and right panels show the evolution of $a(w)/a_I$ and $b(w)/(a_I E)$ respectively for massless vector field domination. Let us note, that $b(0)=0$. In both panels solid blue lines and dashed red lines represent the evolution of scale factors and its low energy limit respectively.}
\label{fig:abBB}
\end{figure}

First of all let us consider a massless vector field domination. This scenario has been briefly analysed in \cite{Artymowski:2010by}. When the energy-stress tensor consist only of the massless vector field, then $\rho_A=\rho=p$. From eq. (\ref{eq:Ein00v},\ref{eq:Ein33v}) one obtains
\begin{equation}
H^2+2H\mathcal{H}=2\frac{\ddot{a}}{a}+H^2\Rightarrow \frac{\ddot{a}}{\dot{a}}=\frac{\dot{b}}{b}\Rightarrow b=E\dot{a}\ , \label{eq:b=aprim}
\end{equation}
where $E=const>0$ has a dimension of time. This shows, that the Universe shrinks along the $z$ axis if only $\ddot{a}<0$. Let us note, that the eq. (\ref{eq:b=aprim}) is also valid in the presence of additional perfect fluid as long as $p_f=-\rho_f$. The equation of motion for a massless vector field looks as follows
\begin{equation}
\dot{\rho}_A+4H\rho_A=0\Rightarrow \rho_A=\rho_{Ao}\left(\frac{a_o}{a}\right)^4\ , \label{eq:rhooda}
\end{equation}
where $\rho_o=\rho(t_o)$ and $t_o$ is any fixed moment in the Universes history. To calculate $a(t)$ let us combine eq. (\ref{eq:Ein33v},\ref{eq:rhooda}) which together give
\begin{equation}
2\ddot{a}a+\dot{a}^2=\rho_oa_o^4a^{-2}\ . \label{eq:ruchua}
\end{equation}
Let us note, that this equation is also valid in the presence of additional perfect fluid with $p_f=0$. The eq. (\ref{eq:ruchua}) can be simplified to
\begin{equation}
\dot{x}=\frac{\sqrt{\rho_I}}{x}\sqrt{x-1}\ ,\label{eq:ruchux}
\end{equation}
where $\rho_I$ is a constant of integration, which has a dimensionality of energy density. The $x(t)$ is defined by $x=a(t)/a_I$, where $\rho(a_I)=\rho_I$. From the eq. (\ref{eq:ruchux}) one can see that $x\geq 1$, so $\rho_I$ is the maximal allowed energy density of the Universe \cite{Artymowski:2010by}. So far we have made no assumptions about $\rho_I$, but in this case we restrict ourselves to consider $\rho_I\ll M_{pl}^4$. Then GR could be the correct theory of gravity in the whole energy range. The exact solution of the eq. (\ref{eq:ruchux}) is
\begin{equation}
x(t)=\left[\frac{2^{1/3}}{(2+w^2+\sqrt{4w^2+w^4})^{1/3}}+\frac{(2+w^2+\sqrt{4w^2+w^4})^{1/3}}{2^{1/3}}-1\right]\ , \label{eq:aodtdokladnie}
\end{equation}
where $w^2=\frac{9}{4}\rho_I(t-t_I)^2$. From eq. (\ref{eq:b=aprim},\ref{eq:aodtdokladnie}) one obtains $R=0$ for any $t$. For $t\to t_I$ one also obtains $R^{\mu\nu}R_{\mu\nu}\to 4\rho_I^2$ and $R^{\mu\nu\alpha\beta}R_{\mu\nu\alpha\beta}\to 20\rho_I^2$. Thus the massless vector field domination gives no initial curvature singularity of the Universe \footnote{The energy conditions for the vector field domination require $\rho>0$, $p>0$ and $\rho\geq|p|$. The massless vector field satisfies them, since $p=\rho=\dot{A}^2/2b^2>0$.}, which together with $a_I\neq 0, H(t_I)=0$ are features of bouncing solutions, like e.g. the Big Bounce. On the other hand from (\ref{eq:b=aprim},\ref{eq:ruchux}) one obtains $\mathcal{H}\rightarrow\infty, b\rightarrow 0$ for $t\rightarrow t_I$, which looks alike the Big Bang scenario. From now on we will set $t_I$ to be equal $0$. When we are very close to the initial value of the energy density one obtains $w\ll 1$, which gives
\begin{equation}
x(t)\simeq (1+\frac{1}{9}w^2)=(1+\frac{1}{4}\rho_It^2)\ , \qquad b\propto t \ ,\qquad V\propto t\left(1+\frac{1}{4}\rho_It^2\right)^2\ . \label{eq:a,bodtdlaw<<1}
\end{equation}
The Universe expands along the $z$ direction for $w\in (0,4)$. For $\rho\ll\rho_I$ one obtains $w\gg 1$. Then
\begin{equation}
x(t)\simeq  w^{2/3}=\left(\frac{9}{4}\rho_I\right)^{1/3}t^{2/3}\ ,\qquad b\propto t^{-1/3}\propto a^{-1/2} \ ,\qquad V\propto t\ . \label{eq:a,bodtdlaw>>1}
\end{equation}
The low energy limit bring us to the Kasner-like solution of diagonal Bianchi I model, with one direction shrinking and two expanding. Such predictions are obviously excluded by the astronomical observations, but in section \ref{sec:BBreal} we will consider realistic extensions of this model. The evolution of $a(w)$ and $b(w)$ is shown in the fig. \ref{fig:abBB}.

\subsection{Vector field non-minimally coupled to gravity}\label{sec:vecnonminimal}

Let us investigate the case of a vector field with non-minimal coupling to gravity. This kind of coupling has been introduced to provide the slow-roll evolution during inflation for the physical vector field defined by $V(t)=A(t)/b$ \cite{Golovnev:2008cf,Dimopoulos:2006ms,Turner:1987bw}. In such a case a vector field may survive inflation and dominate the Universe after reheating. The action for a non-minimally coupled background vector field looks as follows \cite{Dimopoulos:2010nq}
\begin{equation}
S_A=\int d^4x\sqrt{|g|}\left[\frac{1}{2}R+\frac{1}{2b^2}\dot{A}^2-\frac{1}{2b^2}(m^2+\xi R)A^2\right],\qquad S=S_A+S_f,\label{eq:dzialanieAR}
\end{equation}
where $g=\det(g_{\mu\nu})$. Let us assume, that the perfect fluid is minimally coupled to gravity. Then one obtains the effective Einstein equation in the form
\begin{eqnarray}
H(H+2\mathcal{H})=\rho_f+\frac{\dot{A}^2}{2 b^2}+\frac{2\xi }{b^2}(\mathcal{H}+2 H) A\dot{A}+\left(m^2+2\xi\left(-2 \mathcal{H}^2-2 \mathcal{H} H+H^2\right)\right) \frac{A^2}{2 b^2}, \label{eq:Eineff00AR}\\
\frac{\ddot{a}}{a}+\frac{\ddot{b}}{b}+H\mathcal{H}=-p_\perp=-p_f-(1-4 \xi ) \frac{\dot{A}^2}{2 b^2}-\frac{2\xi }{b^2} A \left((3 \mathcal{H}-H) \dot{A}-\ddot{A}\right)+\nonumber\\
\frac{A^2}{2 b^2} \left(m^2+2\xi  \left(3 \mathcal{H}^2-\mathcal{H} H+H^2-\dot{\mathcal{H}}+\dot{H}\right)\right)\ , \label{eq:Eineff11AR}\\
2\frac{\ddot{a}}{a}+H^2=-p_\parallel=-p_f+(1+4\xi)\frac{\dot{A}^2}{2b^2}-\frac{2\xi }{b^2}A\left((4 \mathcal{H}-2H)\dot{A}-\ddot{A}\right)-\nonumber\\
\frac{A^2}{2b^2} \left(m^2+2\xi\left(-2\mathcal{H}^2+8\mathcal{H}H+3H^2+4  \dot{\mathcal{H}}+2\dot{H}\right)\right)\ . \label{eq:Eineff33AR}
\end{eqnarray}
The bouncing solution appears only if initially energy of a vector field's kinetic term dominates, since we want to avoid $b(t)$ dependence of the parallel pressure and the energy density. From eq. (\ref{eq:Eineff00AR},\ref{eq:Eineff33AR}) for $\dot{A}^2\gg \max\{m^2A^2,RA^2\}$ one obtains approximate equations of motions of scale factors
\begin{equation}
\dot{x}\simeq\frac{1}{x}\sqrt{(1+4\xi)\rho_I}\sqrt{x-1}\ ,\qquad \frac{\ddot{x}}{x}-\frac{\dot{b}}{b}\frac{\dot{x}}{x}\simeq 2\xi\rho_Ix^{-4}\ . \label{eq:ruchuxbnonminimal}
\end{equation}
Thus, from the eq. (\ref{eq:ruchux}) one recovers (\ref{eq:aodtdokladnie}) as a solution for $x(w)$ around bounce, but the definition of $w$ variable is changed to $w^2=\frac{9}{4}\rho_I(1+4\xi)t^2$. We have obtained bouncing solution for $x$ with slightly changed effective time variable, but we need to assume, that $1+4\xi>0$ to keep $w$ real. Let us note, that the non-minimal coupling violates the eq. (\ref{eq:b=aprim}). To find the evolution of $b$ for energies close to $\rho_I$ let us consider $b=b(x)$. Then from the eq. (\ref{eq:ruchuxbnonminimal}) one obtains
\begin{equation}
b\propto (x-1)^{\frac{1}{2(1+4\xi)}}x^{-\frac{1+2\xi}{1+4\xi}}\to 0 \quad\text{for}\quad x\to 1\ ,
\end{equation}
so one recovers all features of bounce generated by a minimally coupled massless vector field.

\subsection{Vector field with non-canonical kinetic term}\label{sec:vecnoncanonical}

Another way to obtain the slow-roll evolution of $\rho_A$ during inflation is to consider a vector field with a non-canonical kinetic term. Such a field may dominate the Universe after inflaton and become the vector curvaton \cite{Dimopoulos:2009vu,Dimopoulos:2009am}. Let us define its action as
\begin{equation}
S=\int d^4x\sqrt{-g}\left(\frac{1}{2}R-f\frac{1}{4}F_{\mu\nu}F^{\mu\nu}+\frac{1}{2}m^2A^\mu A_\mu\right)\ , \label{eq:dzialanieAf}
\end{equation}
where $f=f(a,b), m=m(a,b)$. Time dependence of the mass term is necessary to obtain a constant value of $\mathcal{M}=m/\sqrt{f}$, which is a mass term of a canonically normalised vector field $U=\sqrt{f}A/b$. Einstein equations, together with the equation of motion of $U$, look as follows
\begin{eqnarray}
H(H+2\mathcal{H})=\rho_f+\rho_A=\rho_f+\frac{1}{2b^2}\left(f\dot{A}^2+m^2A^2\right)\ ,\label{eq:Ein00nkanon}\\
\frac{\ddot{a}}{a}+\frac{\ddot{b}}{b}+\mathcal{H}H=-p_\perp=-p_f+\frac{A^2}{2b^2}\left(m^2+\frac{a}{2}m^2_{,a}\right)-\frac{\dot{A}^2}{2b^2}\left(f+\frac{a}{2}f_{,a}\right)\ ,\label{eq:Ein11nkanon}\\
2\frac{\ddot{a}}{a}+H^2=-p_\parallel=-p_f-\frac{\dot{A}^2}{2b^2}\left(f-bf_{,b}\right)+\frac{A^2}{2b^2}\left(m^2-b\ m^2_{,b}\right)\ ,\label{eq:Ein33nkanon}\\
\ddot{U}+(2H+\mathcal{H})\dot{U} +\left(\mathcal{M}^2+2\mathcal{H}H+\dot{\mathcal{H}}+\frac{1}{4}\frac{\dot{f}^2}{f^2}-\frac{1}{2}\frac{\ddot{f}}{f}-\frac{1}{2}(2H-\mathcal{H})\frac{\dot{f}}{f}\right)U=0. \label{eq:ruchuUf}
\end{eqnarray}
To obtain cancellation of all time dependent mass terms in the eq. (\ref{eq:ruchuUf}) one needs $f\propto b^2,m\propto b$. For $f\propto a^{-4},m\propto a^{-2}$ the time dependent part of the mass term is equal to $2\dot{H}+\dot{\mathcal{H}}$, so it is negligible during inflation. Let us note, that for both sets of $f,m$ functions one obtains $p_\perp=p_\parallel$, so $T^\mu_{\ \nu}$ is fully isotropic \footnote{In \cite{Dimopoulos:2009vu,Dimopoulos:2009am} Dimopoulos et al consider $m\propto a$ and $f\propto a^{-4}$ or $f\propto a^2$ to obtain flat power spectra of initial inhomogeneities.}. The continuity equation for a vector field is of the form of
\begin{equation}
\dot{\rho}_A+(2H+\mathcal{H})\left(\rho_A+p_A\right)=0\quad\text{for}\quad f\propto b^2\ ,\quad\dot{\rho}_A+(2H+\mathcal{H})\left(\rho_A-p_A\right)=0\quad\text{for}\quad f\propto a^{-4}\ .\label{eq:ciagloscinkanon}
\end{equation}
One can see, that $f\propto b^2$ gives $T^\mu_{\ \nu}$ identical with a scalar field domination scenario. 

On the other hand, for $f\propto a^{-4}$ one obtains surprising results. For the kinetic term domination one obtains $\rho_A\simeq p_A$, which means that $\dot{\rho}_A\ll H\rho_A$. This case is similar to the slow-roll phase of a scalar field. The potential term domination brings us to $\rho_A\simeq -p_A$, so $\dot{\rho}_A\simeq -6H\rho_A$, which is the evolution similar to the massless scalar field scenario. In general, one can see, that the behaviour of kinetic and potential terms has been swapped, comparing to a massive scalar field scenario. 

For example, for a massless vector field (or massive vector field with kinetic initial conditions) with $f\propto a^{-4}$ one obtains $\dot{\rho}_A=0\Rightarrow\rho_A=const$. Then, from the eq. (\ref{eq:Ein33nkanon}), one obtains
\begin{equation}
2\frac{\ddot{a}}{a}+H^2=\rho_A\Rightarrow\frac{\dot{a}}{a_I}=\dot{x}=\sqrt{\frac{\rho_A}{3x}}\sqrt{x^3-1}\ . \label{eq:ruchuxnkanon}
\end{equation}
We have obtained a bounce for $a=a_I$. From now on we shall replace $\rho_A$ by $\rho_I$, since $\rho_A$ does not need to be constant in the massive case. The solution of the eq. (\ref{eq:ruchuxnkanon}) looks as follows
\begin{equation}
x(t)=\cosh^{2/3}\left[\frac{1}{2} \sqrt{3\rho _I} \left(t-t_I\right)\right]=\cosh^{2/3}\left[\frac{w}{\sqrt{3}}\right]\ ,\label{eq:x(t)nkanon}
\end{equation}
where $w=3\sqrt{\rho_I}(t-t_I)/2$. For a massless vector field, from eq. (\ref{eq:Ein00nkanon},\ref{eq:Ein33nkanon}) one obtains $b\propto\dot{a}\propto\dot{x}$, so we have found analytical solutions for both scale factors. For such an evolution of scale factors one obtains no initial curvature singularity, since $R\to-51\rho_I/4,R^{\mu\nu}R_{\mu\nu}\to 1413\rho_I^2/32,$ and $R^{\mu\nu\alpha\beta}R_{\mu\nu\alpha\beta}\to 549\rho_I^2/16$ for $x\to 1$. Let us note, that for big values of $w$ one obtains
\begin{equation}
a(t)\propto b(t)\propto e^{\sqrt{\rho_I/3}t}\ .
\end{equation}
This exponential expansion is identical with the one of the FRW Universe filled with a cosmological constant. Thus, such a vector field can be a good candidate to play the role of Dark Energy. Let us note, that the Universe becomes isotropic very fast, even without a vector field's mass term \footnote{Or with the negligible influence of the mass term on the evolution of the Universe.}. Let us note, that for $a=b$ the $a(t)$ from the eq. (\ref{eq:x(t)nkanon}) does not satisfy $3H^2=p_I=const$. Thus, anisotropic initial conditions for scale factors are necessary to obtain bouncing solution for $a(t)$. 

On the other hand, one can ask about the naturalness of a kinetic term's domination at early times. For a scalar field it is natural from the point of view of the phase space to consider the kinetic term domination around a bounce. Thus, at early times, due to analogy with scalar fields, one shall expect the domination of a vector field's potential term for $f\propto a^{-4}$. This means, that the bouncing solution is strongly fine tuned if $\mathcal{M}\neq 0$. Let us also note, that the bouncing solution described by the eq. (\ref{eq:x(t)nkanon}) may be also obtained with a cosmological constant. In particular, it may be obtained for a scalar field with potential initial conditions. Thus, such a bouncing solution is not a unique feature of vector fields.

\begin{figure}[h]
\centering
\includegraphics*[height=5cm]{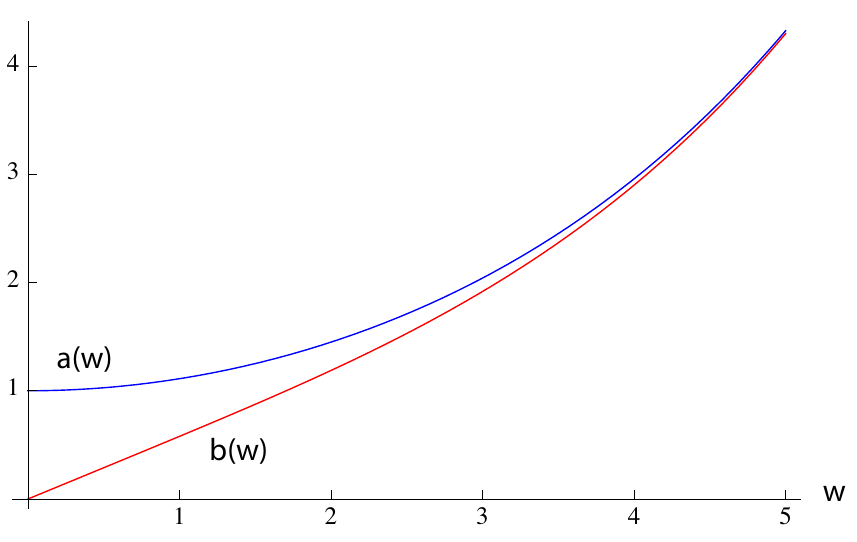}
\hspace{0.5cm}
\includegraphics*[height=5cm]{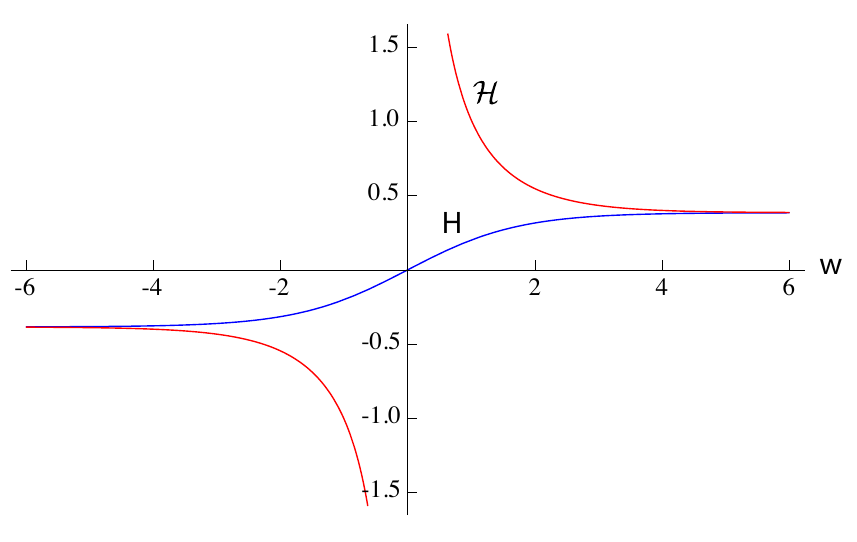}
\caption{The left panel shows the evolution of scale factors for the domination of a massless vector field with non canonical kinetic term and $f\propto a^{-4}$. The right panel presents the evolution of Hubble parameters in the same model. One can see, that after $w\sim t\sqrt{\rho_I}\sim 5$ the Universe becomes isotropic and enters the era of exponential expansion.}
\label{fig:BBnkanon}
\end{figure}

The evolution of scale factors and Hubble parameters for a massless vector field with non-canonical kinetic term is shown in the fig. \ref{fig:BBnkanon}.

\section{Vector field in n+2 dimentions}\label{sec:dominacjaAbezmasnD}

As we have proven, a massless vector field generates a bounce for $a(t)$ and gives the Kasner-like evolution at the low energy limit. Let us consider more generic case of a massless vector field domination in $n+2$ dimensions, where $n$ is a number of dimensions with a rotation symmetry \footnote{This means, that the 4 dimensional space-time is described by $n=2$.}. Let us assume, that the vector field points $x_{n+1}$ axis, which gives the metric tensor of the form of $g_{lk}=\text{Diag}\left(1,-a^2,-a^2,\dots,-a^2,-b^2\right)$, where $l,k=0,1,\dots,n+1$. Then from Einstein equations one obtains
\begin{equation}
\frac{1}{2}n(n-1)H^2+nH\mathcal{H}=n\frac{\ddot{a}}{a}+\frac{1}{2}n(n-1)H^2\Rightarrow \frac{\ddot{a}}{\dot{a}}=\frac{\dot{b}}{b}\Rightarrow b=E\dot{a}\ .\label{eq:b=aprimnDim}
\end{equation}
It is worth to note, that the eq. (\ref{eq:b=aprimnDim}) is valid for any $n$. From the continuity equation one obtains $\rho\propto a^{-2n}$, so from $G^{n+1}_{\ n+1}=p_A$ one obtains the equation of motion for a scale factor $a(t)$
\begin{equation}
\ddot{x}x+\frac{1}{2}(n-1)\dot{x}^2=\frac{\rho_I}{n}x^{2(n-1)}\ ,\label{eq:ruchuanDim}
\end{equation}
where $x(t)=a(t)/a_I$, $a_I=a(t_I)$ and $t_I$ is the moment of a bounce. Let us assume, that $n\neq 1$. Then, from the eq. (\ref{eq:ruchuanDim}) one finds
\begin{equation}
\dot{x}=x^{1-n}\sqrt{\frac{2\rho_I}{n(n-1)}}\sqrt{x^{n-1}-1}\ . \label{eq:ruchuxnDim}
\end{equation}
The eq. (\ref{eq:ruchuxnDim}) has the solution of the form of
\begin{equation}
\frac{\sqrt{2(n-1) n}}{(1+n) \sqrt{\rho_I}} x^{(1+n)/2} \, _2F_1\left[\frac{1+n}{2(1-n)},\frac{1}{2},\frac{n-3}{2 (n-1)},x^{1-n}\right]=t-t_I\ , \label{eq:x(t)nDim}
\end{equation}
where $_2F_1$ is a hypergeometric function. Analogously to the 4 dimensional case a bounce appears for $t=t_I$. The $\rho_I$ is a constant of integration and the maximal energy density of the Universe. The eq. (\ref{eq:x(t)nDim}) can be simplified in two limits. Let us consider the evolution of scale factors around a bounce (which means, that $x-1\ll 1$) and in the low energy limit (for which $x\gg 1$). Then from the eq. (\ref{eq:ruchuxnDim}) one obtains
\begin{equation}
x\simeq 1+\frac{\rho_I}{2n}(t-t_I)^2\quad\text{for}\quad x-1\ll 1,\qquad\qquad x\simeq t^{\frac{2}{n+1}}\quad\text{for}\quad x\gg 1\ .
\end{equation}
Those simplified solutions are valid for any $n$. The $n=1$ case gives the equation of motion of the form of
\begin{equation}
\dot{x}=\sqrt{2\rho_I\ln[x(t-t_I)]}\ .
\end{equation}
Thus, for $x=1$ one obtains $\dot{x}=0$. In such a case there is no analytical solution for $x(t)$. 

\subsection*{The initial curvature singularity of the Universe}

To discuss the issue of initial singularity in $n+2$ dimensions let us express $R,R^{\mu\nu}R_{\mu\nu}$ and $R^{\mu\nu\alpha\beta}R_{\mu\nu\alpha\beta}$ us a function of $x$. From eq. (\ref{eq:b=aprimnDim},\ref{eq:ruchuxnDim}) one obtains
\begin{eqnarray}
R&=&\frac{2}{n}(n-2) x^{-2n}\rho_I\ ,\label{eq:RnDim}\\
R^{\mu\nu}R_{\mu\nu}&=&7\frac{4}{n^2}\left(2n^2-3n+2\right) x^{-4n} \rho_I^2\ ,\label{eq:RRnDim} \\
R^{\mu\nu\alpha\beta}R_{\mu\nu\alpha\beta}&=&\frac{4\rho_I^2 x^{-4 n}}{(n-1) n^2} \left(n^2 (n+1) x^{2(n-1)}-4n^2 (2n-1) x^{n-1}+2\left(8n^3-12 n^2+7 n-2\right)\right).\qquad \label{eq:RRRRnDim}
\end{eqnarray}
In the limit of a bounce, i.e. for $x\to 1$, one obtains
\begin{equation}
R\to\frac{2}{n}(n-2) \rho_I\ ,\quad  R^{\mu\nu}R_{\mu\nu}\to\frac{4}{n^2}\left(2n^2-3n+2\right)\rho_I^2\ ,\quad  R^{\mu\nu\alpha\beta}R_{\mu\nu\alpha\beta}\to\frac{4}{n^2}\left(9n^2-10 n+4\right)\rho_I^2\ .\qquad\qquad\label{eq:limRnDim}
\end{equation}
Let us note, that for $n\neq 0$ all invariants mentioned above are finite around a bounce. Thus, for any $n\neq 0$ one obtains no initial curvature singularity of the Universe.

\section{Isotropic models with classical bounce} \label{sec:BBreal}

\subsection{Massive vector field domination} \label{sec:massiveA}

The fact, that a massive vector field with kinetic initial conditions can give bouncing solution for $a(t)$ has been proven in \cite{Artymowski:2010by}. During a bounce the kinetic term dominates over the potential one. Nevertheless, $A(t)$ grows with time, which leads to vector field's oscillations. During that period the background anisotropy $(H-\mathcal{H})/(H+\mathcal{H})$ decreases like $t^{-1}$ and the Universe becomes isotropic. To find more about the evolution of space-time in such a model see \cite{Artymowski:2010by}.

\subsection{Massless vector field and cosmological constant}\label{sec:A+Lambda}

Analytical solutions of the Einstein equation for the massless vector field domination scenario may be obtained also for a massless vector field with a perfect fluid as long as $p_f$ is not a function of $b(t)$. Then both sides of the eq. (\ref{eq:Ein33v}) are functions of $a(t)$ and its time derivatives. The simplest case is a cosmological constant, which gives $\rho_f=-p_f=\rho_\Lambda=-\Lambda=const$. Then, with a certain choose of $t_I$, one can simplify the eq. (\ref{eq:Ein33v}) to
\begin{equation}
\dot{x}(x)=\frac{1}{x}\sqrt{\left(\rho_{AI}-\frac{\rho_\Lambda}{3}\right)x+\frac{\rho_\Lambda}{3}x^4-\rho_{AI}} \ .\label{eq:ruchuxA+Lambda}
\end{equation}
From the eq. (\ref{eq:ruchuxA+Lambda}) one obtains $\dot{x}\to 0$ for $x\to 1$, so $a_I$ is a minimal value of $a(t)$. Thus, one obtains a bounce similar to the one from a massless vector field domination. The $\rho_I=\rho_\Lambda+\rho_{AI}$ is the maximal value of the energy density, since  $\rho_\Lambda=const$ and $\rho_A=\rho_{AI}x^{-4}$. Let us note, that a cosmological constant, together with a massless vector field, has along the $z$ axis the effective equation of state $p=-\rho$. Then, form eq. (\ref{eq:Ein00v},\ref{eq:Ein33v}) one obtains $b=E\dot{a}$, so the eq. (\ref{eq:b=aprim}) is still valid. Thus, there is a clear analogy between cases with and without a cosmological constant. Simply, one expects the cosmological constant to dominate the Universe for energies much lower than $\rho_I$, so it shall not have significant influence on the evolution of space-time around the bounce. To find the low energy limit of that model let us assume, that $x\gg 1$ and $\rho_\Lambda>0$. Then one finds
\begin{equation}
x(t)\propto\sinh^{2/3}\left(\sqrt{\frac{3\rho_\Lambda}{4}}t\right)\to e^{\sqrt{\rho_\Lambda/3}t}\ , \qquad b\propto\dot{x}\to e^{\sqrt{\rho_\Lambda/3}t}\ ,\label{eq:xywykladniczeA+Lambda}
\end{equation}
so for $x\gg 1$ one obtains isotropic Universe even without a vector field's oscillations. 
\\

For the Universe filled with massive vector field with kinetic initial conditions and with a cosmological constant one obtains the following low energy limit for the evolution of space-time: When a vector field starts to oscillate the Universe becomes isotropic. Vector field's oscillations generate particles and radiation, so in the presence of $\Lambda$ one obtains the observed Universe. This realistic low energy limit brings us a chance to consider a situation, in which there is no need for any theory of gravity wider than GR. Let us also note, that a cosmological constant may come from the massless vector field with non-canonical kinetic term (see section \ref{sec:vecnoncanonical}). Thus, one can obtain a bounce with a good low energy evolution of the Universe using only vector fields.

\subsection{Massless vector field and dust} \label{sec:A+dust}

The eq. (\ref{eq:aodtdokladnie}), which is the analytical solution for $a(t)$ obtained for the massless vector field domination, is also valid for the Universe filled with a massless vector field and dust, since dust does not contributes to $p_\parallel$. Thus, the eq. (\ref{eq:ruchux}) remains unchanged. On the other hand, from the continuity equation, one obtains $\rho_A\sim a^{-4},\ \rho_m\sim a^{-2}b^{-1}$, so the energy density is a function of both scale factors. This means, that even if one defines $\rho_{AI}$ such as $\forall_{a}\ \rho_A<\rho_{AI}$, there may be no $\rho_{mI}$ such that $\forall_{a,b}\ \rho_m<\rho_{mI}$, so the total energy density does not need to be finite at all times. Let us note, that the eq. (\ref{eq:b=aprim}) is not valid, since from eq. (\ref{eq:Ein00v},\ref{eq:Ein33v}) one obtains
\begin{equation}
H\mathcal{H}-\frac{\ddot{a}}{a}=\frac{1}{2}\rho_{m}\neq 0\ . \label{eq:bneqaprim}
\end{equation}
This equation may be written as a equation of motion of $b(t)$. Let us define $x=a/a_I$, $y=b/b_o$, $x_o=a_o/a_I$ where $a_I=a(t_I)$, $b_o=b(t_o)$, $t_o\neq t_I$ is a certain moment and $t_I$ is the moment of a bounce of $a(t)$. The $t_o$ needs to be different than $t_I$ since we do not know, if $\rho_m$ is finite at $t=t_I$. From the eq. (\ref{eq:bneqaprim}) one obtains
\begin{equation}
\dot{x}\dot{y}-\ddot{x}y=\rho_{mo}\frac{x_o^2}{2x}.\label{eq:ruchuy}
\end{equation}
From eq. (\ref{eq:ruchux},\ref{eq:ruchuy}) one finds the evolution of $y$ as a function of $x$
\begin{equation}
y(x)=\frac{\rho_{fo}}{3\rho_{AI}}\frac{x_o^2}{x}(x^2+4x-8)+C\frac{1}{x}\sqrt{x-1}\ , \label{eq:yrozw}
\end{equation}
where $C$ is the constant of integration. Thus, for $x\to 1$ one obtains $(x^2+4x-8)/x\to -3$ and $\sqrt{x-1}/x\to 0$, which means, that for $x$ close enough to $1$ one obtains $y<0$. This means, that the minimal allowed value of $x$ need to be larger that 1. Let us define $x_p$ such, as $y(x_p)=0$. Let us note, that $H(x_p)>0$, since $x_p>1$, which together with $\mathcal{H}(x_p)=\infty$ gives $H(H+2\mathcal{H})(x=x_p)=\infty$. This seems to be natural, since $\rho_{m}\propto(x^2y)^{-1}\Rightarrow\rho_{m}(x_p)=\infty$. Thus, by adding dust to a massless vector field one obtains bounce of $a(t)$, which happens in the forbidden region of $y<0$. The beginning of the Universe appears for $y=0$, for which the energy density is infinite. One cannot extend the evolution of the Universe for $t<t_p$, so the considered model has all features of the Big Bang. The only difference is, that the beginning of space-time is not a point, but a circle described by $y=0,x=x_p$.
\\

Similar results can be found for the Universe filled with massless vector field, cosmological constant and dust. There for $x\to 1$ one obtains $y\to-\rho_{m_o}x_o^2/(\rho_{AI}+\Lambda)$, so for $\Lambda<0$ and $\rho_A<-\Lambda$ one obtains $y>0$ for $x<x_p$. Thus, one can solve the problem of negative $y$ at the moment of bounce. However, this generates another problem, since $y$ becomes negative for $x>x_p$. In such a case the Universe would start with a bounce to immediately end up in the Big Crunch. This means, that $y(x)$ may be positive for $x<x_p$ or for $x>x_p$, but one cannot obtain $y>0$ at all times.

\section{Conclusions}\label{sec:concl}

In the initial section of this paper we have shown, that a massless vector field domination generates bouncing evolution of space-time in a plane transverse to a vector field's direction. The bouncing solution us has many features of other bounces known in cosmology, such as a non-zero initial value of the scale factor $a(t)$, lack of initial curvature singularity and finite initial (maximal) energy density. Along the vector field's direction one obtains the evolution of the scale factor and the Hubble parameter similar to the Big Bang model, i.e. $b\to 0$ and $\mathcal{H}\to-\infty$ for $t\to t_I$. The Kasner-like solution with $a\sim t^{2/3},\ b\sim t^{-1/3}$ is the low energy limit of the evolution of space-time.
\\

In section \ref{sec:dominacjaAbezmasnD} we have considered bouncing solution for a massless vector field domination in the general $n+2$ dimensional case, where $n$ is a number of dimensions with isotropic evolution. We have found the analytical solution for the scale factor $a(t)$, as well as its simple form in the high and low energy limit. We have also proven, that $R,R^{\mu\nu}R_{\mu\nu}$ and $R^{\mu\nu\alpha\beta}R_{\mu\nu\alpha\beta}$ do not diverge for $t\to t_I$ for all $n\neq 0$. 
\\

In section \ref{sec:BBreal}, to obtain isotropic low energy limit of the evolution of space-time, we have discussed several extensions of the massless vector field domination scenario. All of them obtain bounce at $(x,y)$ plane for $t=t_I$. First of all we have considered massive vector field with kinetic initial conditions. I such a case the Universe has bouncing solution for $\rho\sim\rho_I$ and isotropic low energy limit, which is generated by the vector field's oscillations. Secondly,l we have considered the Universe filled with a massless vector field and a cosmological constant. In this case a cosmological constant contributes to $\rho_I$ and generates isotropic expansion for $a\gg a_I$. The massive vector field with cosmological constant may give realistic low energy limit with particles and radiation originating from the $A_\mu$ and with a Dark Energy coming from $\Lambda$. We have also considered the Universe filled with massless vector field and dust, for which the bounce appears in the forbidden region $b<0$. The true beginning of space-time appears at $b=0,\mathcal{H}=\infty,a>a_I,H\neq 0$, so the initial energy density diverges and one obtains all features of the Big Bang. Similar results are given by a massless vector field, dust and cosmological constant domination. 

Let us also note, that the presented model has several advantages comparing to bounces from e.g. $f(R)$ theories or ekpyrotic$\backslash$cyclic Universe. GR remains the only theory of gravity needed and the matter sector has the standard form known from particle physics.

\begin{center}
{\bf Acknowledgements}
\end{center}
This work was partially supported by Polish Ministry for Science and Education under grant N N202 091839.

\end{document}